\documentstyle[preprint,aps]{revtex}
\tightenlines
\begin{document}
\draft
\title{TRAJECTORY-COHERENT STATES\\
FOR THE CALDIROLA - KANAI HAMILTONIAN}
\author{A.G. Karavayev}
\address{Tomsk Polytechnic University , 634004 Tomsk, Russia }
\date{\today }
\maketitle

\begin{abstract}
In this paper we construct the trajectory-coherent states of a damped
harmonic oscillator. We investigate the properties of this states.
\end{abstract}


\narrowtext

\section{Introduction}

At present the coherent states are widely used to describe many fields of
theoretical physics\cite{1}. The interest and activity in coherent states
were revived by the paper of Glauber\cite{2}, who showed that the coherent
states could be successfully used for problem of quantum optics.Recently
Nieto and Simmons\cite{3} have constructed coherent states for particles in
general potential,Hartley and Ray\cite{4} have obtained coherent states for
time-dependent harmonic oscillator on the basis of Lewis-Risenfeld
theory;Yeon,Um and George\cite{5} have constructed exact coherent states for
the damped harmonic oscillator.

Some time ago Bagrov,Belov and Ternov\cite{6} have constructed approximate
(for $\hbar \rightarrow {0}$ ) solutions of the Schr\"odinger equation for
particles in general potentials, such that the coordinate and momentum
quantum - mechanical averages were exact solutions of the corresponding
classical Hamiltonian equations; these states were called trajectory-
coherent (TCS). The basis of this construction is the complex WKB method by
V.P.Maslov\cite{7,8,9}.

The aim of this work is to construct the trajectory-coherent (TCS) states of
a damped harmonic oscillator by using the Caldirola - Kanai Hamiltonian and
the complex WKB method. It is shown that this states satisfy the
Schr\"odinger equation exactly.

\section{The construction of TCS.}

Consider the Schr\"{o}dinger equation
\begin{equation}
i\hbar\partial_t\Psi=\hat{H}\Psi,
\end{equation}
where the symbol of operator $\hat{H}$ - the function $H(x,p,t)$ is
arbitrary real and analytical function of coordinate and momentum.The method
of construction TCS in this case has been described in detail in\cite{6},\
hence we shall illustrate some moments only. For constructing the TCS of the
Schr\"{o}dinger equation it is necessary to solve the classical Hamiltonian
system
\begin{equation}
\dot{x}(t)=\partial_p{H(x,p,t)},~ \dot{p}(t)=-\partial_x{H(x,p,t)},
\end{equation}
and the system in variations (this is the linearization of the Hamiltonian
system in the neighbourhood of the trajectory $x(t),p(t)$)
\begin{eqnarray}
\dot{w}(t)=-H_{xp}(t)w(t)-H_{xx}(t)z(t),& w(0)=b, \\
\dot{z}(t)=H_{pp}(t)w(t)+H_{px}(t)z(t), & z(0)=1,  \nonumber
\end{eqnarray}
where $H(x,p,t)$ is the classical Hamiltonian;
\begin{eqnarray}
H_{xp}(t)= \partial_x\partial_p{H(x,p,t)} \mid_{x=x(t),p=p(t)},  \nonumber \\
H_{px}(t)= \partial_p\partial_x{H(x,p,t)} \mid_{x=x(t),p=p(t)},  \nonumber \\
H_{xx}(t)= \partial_{xx}^2 {H(x,p,t)} \mid_{x=x(t),p=p(t)},  \nonumber \\
H_{pp}(t)= \partial_{pp}^2 {H(x,p,t)} \mid_{x=x(t),p=p(t)},  \nonumber
\end{eqnarray}
$b$ is complex number obeying the condition ${\bf Im}b> 0$,~and ~$x(t),~p(t)
$~are the solutions of system (2).

Consider the damped harmonic oscillator\cite{5,10,11}
\begin{equation}
H(x,p,t)=\exp(-\gamma t)(2m)^{-1}p^2+\frac{1}{2}\exp(\gamma t)m \omega_0^2
x^2.
\end{equation}
The Lagrangian and mechanical energy are given by\cite{5}
\begin{eqnarray}
L=\exp(\gamma t)(\frac {1}{2}m\dot{x}^2-\frac {1}{2}m\omega_0^2 x^2), \\
E=\exp(-2\gamma t)(2m)^{-1}p^2+\frac{1}{2}m\omega_0^2 x^2.  \nonumber
\end{eqnarray}
We first define the Cauchy problem for equation (1)
\begin{eqnarray}
\mid 0>\mid_{t=0}=\Psi_0(x,t,\hbar)\mid_{t=0}= N\exp\{i\hbar^{-1}(p_0(x-x_0)
\\
+\frac{b}{2}(x-x_0)^2)\},  \nonumber
\end{eqnarray}
where $x_0=x(t)\mid_{t=0},~p_0=p(t)\mid_{t=0}.$

The function of WKB - solution type (TCS)\cite{6}
\begin{equation}
\mid 0>=\Psi _0(x,t,\hbar )=N\Phi (t)\exp \{i\hbar ^{-1}S(x,t)\},
\end{equation}
\ where $N=({\bf Im}b(\pi \hbar )^{-1})^{1/4},~\Phi (t)=(z(t))^{-1/2},$
\begin{eqnarray}
S(x,t) &=&\int\limits_0^t{\{\dot x(t)p(t)-H(x(t),p(t),t)\}dt}  \nonumber \\
&&+p(t)(x-x(t))+\frac 12w(t)z^{-1}(t)(x-x(t))^2,  \nonumber
\end{eqnarray}
and the phase $S(x,t)$ is the complex - valued function $({\bf Im}S>0)$ is
the approximate solution of the Cauchy problem (6) for the Schr\"odinger
equation (1). We should note that in the case of the quadratic systems, for
example, for Hamiltonian (4) the function (7) is the exact solution of the
equation (1).

Solving the differential equations (2),(3) we obtain
\begin{equation}
x(t)=\frac 1{2m\omega }\exp (-\frac 12\gamma t)\left( 2p_0\sin \omega
t+m(2\omega \cos \omega t+\gamma
\mathop{\rm sin}
\omega t)x_0\right) ,
\end{equation}
\[
p(t)=-\frac 1{4\omega }\exp (\frac 12\gamma t)\left( (2\gamma
\mathop{\rm sin}
\omega t-4\omega \cos \omega t)p_0+m(\gamma ^2+4\omega ^2)x_0%
\mathop{\rm sin}
\omega t\right) ,
\]
\[
w(t)=-\frac 1{4\omega }\exp (\frac 12\gamma t)\left( (2b\gamma +m(\gamma
^2+4\omega ^2))%
\mathop{\rm sin}
\omega t-4b\omega \cos \omega t\right) ,
\]
\[
z(t)=\exp (-\frac 12\gamma t)\left( \cos \omega t+\frac{(2b+m\gamma )\sin
\omega t}{2m\omega }\right) ,
\]
\[
~\omega ^2=\omega _0^2-\frac 14\gamma ^2.
\]
It is easy to check that the function (7), where~$x(t),p(t),w(t),z(t)$~ are
defined by (8) satisfy the equation (1) exactly as for ~$\omega ^2>0$~as ~$%
\omega ^2<0~~(\omega \rightarrow i\omega )$~.

Further,we define ''annihilation'' operator~ $\hat a(t)$~ and ''creation''
operator ~$\hat a^{+}(t)$ as\cite{6}:
\begin{equation}
\hat a(t)=(2\hbar {\bf Im}b)^{-1/2}\{z(t)(\hat p-p(t))-w(t)(x-x(t))\},
\end{equation}
\[
\hat a^{+}(t)=(2\hbar {\bf Im}b)^{-1/2}\{z^{*}(t)(\hat p%
-p(t))-w^{*}(t)(x-x(t))\}.
\]
It easy to check that the creation and annihilation operators satisfy the
usual Bose permutation rule
\begin{equation}
\lbrack \hat a,\hat a^{+}]=1,~[\hat a,\hat a]=[\hat a^{+},\hat a^{+}]=0.
\end{equation}
The complete orthonormal set of the trajectory-coherent states (TCS) are
defined as\cite{6}:
\begin{equation}
\mid n>=(n!)^{-1/2}(\hat a^{+})^n\mid 0>.
\end{equation}
It is not difficult to check the relations
\begin{equation}
<n\mid m>=\delta _{n,m},~~~~\hat a\mid 0>=0, \\
\end{equation}
\[
\hat a^{+}\mid n>=(n+1)^{1/2}\mid n+1>,~~~\hat a\mid n>=(n)^{1/2}\mid n-1>.
\]

Using the usual methods\cite{1} we obtain the expression for coherent states
(CS)
\begin{equation}
\mid \alpha >=\hat{A}(\alpha )\mid 0>,
\end{equation}
where $\hat{A}(\alpha )$ is the unitary operator
\begin{equation}
\hat{A}(\alpha )=\exp (\alpha \hat{a}^+-\alpha ^* \hat{a}),
\end{equation}
~$\alpha $~ is the complex number;~ $\hat{a}^+, \hat{a}$~ are defined by (9).

Besides, at is follows from (12)-(14), the functions~ $\mid~\alpha >$~ are
eigenstates of the operator~ $\hat{a}$~ with eigenvalue~ $\alpha $,\ i.e.
\begin{equation}
\hat{a} \mid \alpha >=\alpha \mid \alpha >.
\end{equation}

\section{Quantum - mechanical averages and uncertainty relations.}

Further we shall find the expressions for quantum - mechanical averages 
\[
<\hat{x}>_{\scriptscriptstyle TCS},\ <\hat{x}^2>_{\scriptscriptstyle TCS},\ <%
\hat{p}>_{\scriptscriptstyle TCS},\ <\hat{p}^2>_{\scriptscriptstyle TCS},\ <%
\hat{x}>_{\scriptscriptstyle CS},\ 
\]
\[
<\hat{x}^2>_{\scriptscriptstyle CS}, <\hat{p}>_{\scriptscriptstyle CS},\ <%
\hat{p}^2>_{\scriptscriptstyle CS},\ <\hat{E}>_{\scriptscriptstyle TCS},\ <%
\hat{E}>_{\scriptscriptstyle CS},\ 
\]
where, for example, 
\[
<\hat{x}>_{\scriptscriptstyle TCS}=<n\mid \hat{x} \mid n>,~ <\hat{x}>_{%
\scriptscriptstyle CS}=<\alpha \mid \hat{x} \mid \alpha>. 
\]

Further,we present the relations expressing the coordinate and momentum
operators in terms of the operators ~ $\hat{a}^+, \hat{a}$~ 
\begin{equation}
\hat{x}=x(t)-i(2{\bf Im}b(\hbar)^{-1})^{-1/2}\{z(t)\hat{a}^+-z^*(t)\hat{a}\},
\end{equation}
\[
\hat{p}=p(t)-i(2{\bf Im}b(\hbar)^{-1})^{-1/2}\{w(t)\hat{a}^+-w^*(t)\hat{a}%
\}. 
\]
Using (5),(9)-(16), we obtain for quantum - mechanical averages: 
\begin{equation}
<\hat{x}>_{\scriptscriptstyle TCS}=x(t),~ <\hat{p}>_{\scriptscriptstyle %
TCS}=p(t), \\
\end{equation}
\[
<\hat{x}>_{\scriptscriptstyle CS}=x(t)-i(2{\bf Im}b(\hbar)^{-1})^{-1/2}\{
\alpha^* z(t)- \alpha z^*(t)\}, 
\]
\[
<\hat{p}>_{\scriptscriptstyle CS}=p(t)-i(2{\bf Im}b(\hbar)^{-1})^{-1/2}\{%
\alpha^* w(t)-\alpha w^*(t)\}, 
\]
\[
<\hat{x}^2>_{\scriptscriptstyle TCS}= x^2(t)+\frac{\hbar}{{\bf Im}b}(n+\frac{%
1}{2})|{z(t)}|^2, 
\]
\[
<\hat{p}^2>_{\scriptscriptstyle TCS}= p^2(t)+\frac{\hbar}{{\bf Im}b}(n+\frac{%
1}{2})|{w(t)}|^2, 
\]
\[
<\hat{x}^2>_{\scriptscriptstyle CS}= <x>_{\scriptscriptstyle CS}^2+\frac{%
\hbar}{2{\bf Im}b}|{z(t)}|^2, 
\]
\[
<\hat{p}^2>_{\scriptscriptstyle CS}= <p>_{\scriptscriptstyle CS}^2+\frac{%
\hbar}{2{\bf Im}b}|{w(t)}|^2. 
\]
\begin{eqnarray*}
<\hat{E}>_{\scriptscriptstyle TCS}= \exp(-2\gamma t)(2m)^{-1}p^2(t) \\
+\frac{1}{2}m\omega_0^2 x^2(t)+\frac{\hbar}{{\bf Im}b}(n+\frac{1}{2}) \\
\times \{\exp(-2\gamma t)(2m)^{-1}|{w(t)}|^2 +\frac{1}{2}m\omega_0^2|{z(t)}%
|^2\},
\end{eqnarray*}
\begin{eqnarray*}
<\hat{E}>_{\scriptscriptstyle CS}= \exp(-2\gamma t)(2m)^{-1}<p>_{%
\scriptscriptstyle CS}^2 \\
+\frac{1}{2}m\omega_0^2<x>_{\scriptscriptstyle CS}^2 +\frac{\hbar}{2{\bf Im}b%
} \\
\times \{\exp(-2\gamma t)(2m)^{-1}|{w(t)}|^2 +\frac{1}{2}m\omega_0^2|{z(t)}%
|^2\}.
\end{eqnarray*}
Now, by calculating the uncertainty in\ $x$\ and\ $\hat{p}$\ in the TCS and
CS one finds: 
\begin{eqnarray}
(\Delta\hat{x})_{\scriptscriptstyle TCS}^2= <(\hat{x}-<\hat{x}>_{%
\scriptscriptstyle TCS})^2>_{\scriptscriptstyle TCS} \\
=\hbar(n+\frac{1}{2})\frac{|{z(t)}|^2}{{\bf Im}b},  \nonumber \\
(\Delta\hat{p})_{\scriptscriptstyle TCS}^2= <(\hat{p}-<\hat{p}>_{%
\scriptscriptstyle TCS})^2>_{\scriptscriptstyle TCS}  \nonumber \\
=\hbar(n+\frac{1}{2})\frac{|{w(t)}|^2}{{\bf Im}b},  \nonumber
\end{eqnarray}
\begin{eqnarray}
(\Delta\hat{x})_{\scriptscriptstyle CS}^2= <(\hat{x}-<\hat{x}>_{%
\scriptscriptstyle CS})^2>_{\scriptscriptstyle CS} = \frac{\hbar}{2} \frac{|{%
z(t)}|^2}{{\bf Im}b}, \\
(\Delta\hat{p})_{\scriptscriptstyle CS}^2= <(\hat{p}-<\hat{p}>_{%
\scriptscriptstyle CS})^2>_{\scriptscriptstyle CS} = \frac{\hbar}{2} \frac{|{%
w(t)}|^2}{{\bf Im}b}.  \nonumber
\end{eqnarray}
So,\ the Heisenberg's uncertainty relations is expressed as 
\begin{equation}
(\Delta\hat{x})_{\scriptscriptstyle TCS}^2 (\Delta\hat{p})_{%
\scriptscriptstyle TCS}^2= \hbar^2 (n+\frac{1}{2})^2 \frac{| w(t)z(t) | ^2}{(%
{\bf Im}b)^2},
\end{equation}
\begin{equation}
(\Delta\hat{x})_{\scriptscriptstyle CS}^2 (\Delta\hat{p})_{%
\scriptscriptstyle CS}^2= \frac{\hbar^2}{4} \frac{| w(t)z(t) | ^2}{({\bf Im}%
b)^2}.
\end{equation}

For the damped harmonic oscillator (4) we choose\ ${\bf Re}b=0$\ (it is
necessary for minimization of the uncertainty relations in the initial
instant of time\cite{12}),\ ${\bf Im}b=\mu m\omega $\ , where\ $\mu >0$\
shows the initial uncertainty of coordinate;\ and we denote\ $\theta =\gamma
/{2\omega }$.

By using (8),(16)-(21), in the case~\ $\omega^2=\omega_0^2-\frac{1}{4}%
\gamma^2>0$~ for\ \ $(\Delta\hat{x})_ {\scriptscriptstyle TCS}^2,\\(\Delta%
\hat{p})_ {\scriptscriptstyle TCS}^2 ,\ (\Delta\hat{x})_ {\scriptscriptstyle %
CS}^2,\ (\Delta\hat{p})_ {\scriptscriptstyle CS}^2$\ \ we obtain 
\begin{eqnarray*}
(\Delta\hat{x})_{\scriptscriptstyle TCS}^2 =\hbar(n+\frac{1}{2})\exp(-\gamma
t) (\mu m \omega)^{-1} \{1+\sin^2\omega t(\theta^2 \\
+\mu^2-1)+\theta \sin2\omega t\},
\end{eqnarray*}
\begin{eqnarray*}
(\Delta\hat{p})_{\scriptscriptstyle TCS}^2 =\hbar(n+\frac{1}{2})\exp(\gamma
t)\mu m \omega\{1+\frac{1}{\mu^2}\sin^2\omega t(2\theta^2 \\
+\theta^4+1+\mu^2\theta^2-\mu^2)-\theta\sin2\omega t\},
\end{eqnarray*}
\begin{eqnarray*}
(\Delta\hat{x})_{\scriptscriptstyle CS}^2 =\frac{\hbar}{2}\exp(-\gamma t)
(\mu m \omega)^{-1} \{1+\sin^2\omega t(\theta^2 \\
+\mu^2-1)+\theta \sin2\omega t\},
\end{eqnarray*}
\begin{eqnarray*}
(\Delta\hat{p})_{\scriptscriptstyle CS}^2 =\frac{\hbar}{2}\exp(\gamma t)\mu
m \omega\{1+\frac{1}{\mu^2}\sin^2\omega t(2\theta^2 \\
+\theta^4+1+\mu^2\theta^2-\mu^2)-\theta\sin2\omega t\},
\end{eqnarray*}
and thus the uncertainty relations becomes 
\begin{equation}
(\Delta\hat{x})_{\scriptscriptstyle TCS}^2 (\Delta\hat{p})_{%
\scriptscriptstyle TCS}^2 =\hbar^2(n+\frac{1}{2})^2(1+g(t)),
\end{equation}
\[
(\Delta\hat{x})_{\scriptscriptstyle CS}^2 (\Delta\hat{p})_{%
\scriptscriptstyle CS}^2 =\frac{\hbar^2}{4}(1+g(t)), 
\]
where 
\begin{eqnarray}
g(t)=\{\frac{\theta}{\mu}(\theta^2+\mu^2+1)\sin^2\omega t \\
+\frac{1}{2\mu}(\theta^2-\mu^2+1)\sin2\omega t\}^2.  \nonumber
\end{eqnarray}
We should note,that in the special case\ $\mu =1$\ the formula (23)
coincides with formula (14) from\cite{5}. The function (23) is equal to
zero,and,therefore, the minimization of the uncertainty relations has place
in the instants of time 
\begin{eqnarray*}
t_{1k}=\frac{\pi k}{\omega},~~~t_{2k}=\frac{1}{\omega} \arctan \frac{%
\mu^2-\theta^2-1}{\theta(\mu^2+\theta^2+1)} \\
+\frac{\pi k}{\omega};\ ~ k=0,1,2,\ldots .
\end{eqnarray*}
The preceding equation can be solved for the parameter\ $\mu >0$\ in the
case 
\begin{equation}
|\theta \tan \omega t|<1.
\end{equation}
and,therefore,by choosing parameter\ $\mu $\ we can obtain the minimization
for any instant of time\ $t$\ obeying the condition (24).

In the case\ $\omega ^2=\frac 14\gamma ^2-\omega _0^2>0$\ we obtain 
\begin{eqnarray*}
(\Delta \hat x)_{\scriptscriptstyle TCS}^2 &=&\hbar (n+\frac 12)\exp
(-\gamma t)(\mu m\omega )^{-1} \\
&&\ \times \{1+\sinh ^2\omega t(\theta ^2+\mu ^2+1)+\theta \sinh 2\omega t\},
\end{eqnarray*}
\begin{eqnarray*}
(\Delta \hat p)_{\scriptscriptstyle TCS}^2 &=&\hbar (n+\frac 12)\exp (\gamma
t)\mu m\omega \\
&&\ \times \{1+\frac 1{\mu ^2}\sinh ^2\omega t(1-2\theta ^2+\theta ^4+\mu
^2\theta ^2+\mu ^2) \\
&&\ -\theta \sinh 2\omega t\},
\end{eqnarray*}
\begin{eqnarray*}
(\Delta \hat x)_{\scriptscriptstyle CS}^2 &=&\frac \hbar 2\exp (-\gamma
t)(\mu m\omega )^{-1}\{1+\sinh ^2\omega t(\theta ^2 \\
&&\ +\mu ^2+1)+\theta \sinh 2\omega t\},
\end{eqnarray*}
\begin{eqnarray*}
(\Delta \hat p)_{\scriptscriptstyle CS}^2 &=&\frac \hbar 2\exp (\gamma t)\mu
m\omega \{1+\frac 1{\mu ^2}\sinh ^2\omega t(1-2\theta ^2 \\
&&\ +\theta ^4+\mu ^2\theta ^2+\mu ^2)-\theta \sinh 2\omega t\},
\end{eqnarray*}
and for the uncertainty relations we have 
\[
(\Delta \hat x)_{\scriptscriptstyle TCS}^2(\Delta \hat p)_{%
\scriptscriptstyle TCS}^2=\hbar ^2(n+\frac 12)^2(1+g(t)), 
\]
\[
(\Delta \hat x)_{\scriptscriptstyle CS}^2(\Delta \hat p)_{\scriptscriptstyle %
CS}^2=\frac{\hbar ^2}4(1+g(t)), 
\]
where 
\begin{eqnarray}
g(t) &=&\{\frac \theta \mu (\theta ^2+\mu ^2-1)\sinh ^2\omega t \\
&&\ +\frac 1{2\mu }(\theta ^2-\mu ^2-1)\sinh 2\omega t\}^2.  \nonumber
\end{eqnarray}
Now the function\ $g(t)$\ (25) is equal to zero only for 
\begin{eqnarray*}
~~t_{01}=0~~,~~t_{02}=\frac 1\omega {\rm arctanh}\frac{\mu ^2-\theta ^2+1}{%
\theta (\mu ^2+\theta ^2-1)}.
\end{eqnarray*}
This equation can be solved for parameter \ $\mu $\ in the case 
\begin{equation}
|\theta \tanh \omega t|<1\ (\theta >1),\qquad |\theta \tanh \omega t|>1\
(0<\theta <1)
\end{equation}
and we can obtain the minimization for any instant of time\ $t$\ obeying the
condition (26).We should notice, that the formula (23) can be easily
obtained from (25) by means of replacement $\omega \rightarrow i\omega $
and, therefore $\mu \rightarrow -i\mu ,\theta \rightarrow -i\theta .$ In a
case of harmonic oscillator $(\gamma =0)$ as (23) and (25) bring to $g(t)=0$%
, that coincides with results, received in Ref.12.

\section*{Acknowledgments}

We thank Prof.\ A.Yu.Trifonov and Prof.\ D.V.Boltovsky for useful
discussions.

\end{document}